\documentclass{elsart}
\usepackage{graphicx}
\usepackage{amssymb}

\begin{document}

\begin{frontmatter}

\title{Power Law in Firms Bankruptcy}

\author[hong]{Byoung Hee Hong}
\author[jwl]{Kyoung Eun Lee}
\author[jwl]{Jae Woo Lee\corauthref{cor1}}
\ead{jaewlee@inha.ac.kr}
\corauth[cor1]{Corresponding author. Tel., 82+32+8607660; fax, 82+32+8727562}
\address[hong]{Department of Electrophysics, Kwangwoon University, Seoul 139-701, Korea}
\address[jwl]{Department of Physics, Inha University, Incheon 402-751, Korea}

\begin{abstract}
We consider the scaling behaviors for 
fluctuations of the number of Korean firms bankrupted
in the period from August 1 2002 to October 28 2003. 
We observe a power law for the distribution of the number of the
bankrupted firms. The Pareto exponent is close to unity.
We also consider the daily increments of the number 
of firms bankrupted. The probability distribution of the daily increments for the firms bankrupted
follows the Gaussian distribution in central part and has a fat tail.
The tail parts of the probability distribution of the daily increments
for the firms bankrupted follow a power law. 
\end{abstract}


\end{frontmatter}

\newcommand{\be}{\begin{equation}}
\newcommand{\ee}{\end{equation}}

Zipf law is appeared in many natural and social systems such as the distribution
of city sizes, the size of earthquakes, moon craters, solar flares, the intensity
of wars, the frequency of use in any human language, the sales of books, 
the number of species in biological 
taxa\cite{SO03,ZI49,GR44,GE94,LH91,RT98,CF95,WY22}.
In Zipf law or Pareto law, the distribution or histogram follows
a power law, $p(x) \sim x^{-(\alpha+1)}$ where $\alpha$ is a Pareto exponent. 

Zipf law has also been reported in economic systems such as the income distribution of
companies, the sales of books and almost every other bounded commodity,
the distribution of bank assets,
 and the distribution of firm's debt when the company
is bankrupted\cite{FU04,OT99,AS00,IS05,PA04,Lee1,Lee2,Lee3}. 
Fujiwara has reported the Zipf law in firms bankrupted in Japan. He found
a power law, $P(>x) \sim x^{-\alpha}$ with $\alpha=0.911(8)$
 for the cumulative distribution of firm's debt\cite{FU04}. 
Okuyama and Takayasu reported Zipf law with a Pareto exponent $-1$ 
for the income distribution of Japanese companies\cite{OT99}.
Aoyama et al. analyzed the distribution of income and income tax of
individuals in Japan. They observed the power law with the Pareto exponent
$-2$\cite{AS00}.
Ishkawa reported the income
distribution in Japan. He observed Pareto law with the Pareto index $-1$
 for the income distribution of companies\cite{IS05}.

We observe the power law for the distribution of the number of firms bankrupted.
The Pareto exponent is close to unity. We also observe the power law for daily
increments of the firms bankrupted. 

\begin{figure}[b!]
\begin{center}
\includegraphics[width=8cm,height=10cm,angle=270,clip]{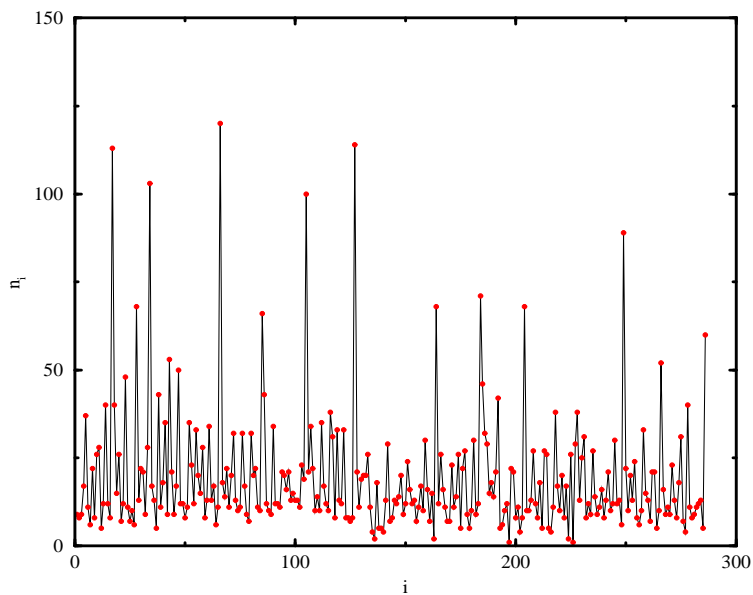}
\end{center}
\caption{The daily number of the firms bankrupted versus the time (days)
in Korea from August 1 2002 to October 28 2003.}
\label{f1}
\end{figure}

\begin{figure}[t]
\begin{center}
\includegraphics[width=8cm,height=10cm,angle=270,clip]{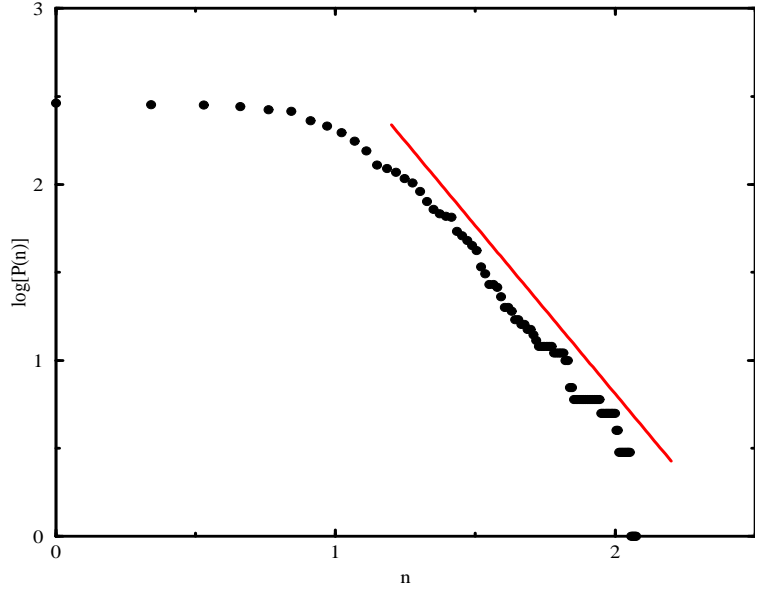}
\end{center}
\caption{The log-log plot of the cumulative probability distribution $P(x>n)$
for the number of the firms bankrupted versus the time (days). 
The solid line is the least-square fit with a slope $-0.91(2)$.}
\label{f2}
\end{figure}


We consider the firms bankrupted in Korea in the period from August 1 2002 
to October 28 2003\cite{RA}.
The total number of the recorded date are 286 days. We deleted the weekends and holidays
from the time series. We count the number of daily bankrupted firms.
Let $n_i$ be the number of the firms bankrupted at date $i$.
Fig. 1 presents the number $n_i$ of the firms bankrupted as a function of the date $i$.
The time series of the firms bankrupted show many big peaks.
We can observe
that the number of the bankruptcies is increased before the big crash.
The average number of the firms bankrupted is $<n_i > =19.3 $ and the standard
deviation is $ \sigma_n =17.8$. 

Consider the probability distribution $p(n)$ for the number of the firms bankrupted. We expect
the power law at the large $n$,
\begin{equation}
p(n) \sim n^{-(\alpha+1)}.
\end{equation}
When the probability distribution $p(n)$ follows the power law, the cumulative probability
distribution also shows the power law as
\begin{equation}
P(x>n) \sim n^{-\alpha},
\end{equation}
where $\alpha$ is a Pareto index.

Fig. 2 presents a log-log plot of the cumulative probability distribution $P(x>n)$ 
for the number of the firms bankrupted. We observe the obvious power law 
and obtain the Pareto index $\alpha=0.91(2)$ by a least-square fit.
The number of the firms bankrupted shows the Zipf law.
This Pareto index is compared to the Pareto index $\alpha=0.911(8)$ 
for the distribution of firm's debt when the firm was bankrupted in Japan\cite{FU04}.

\begin{figure}[tb]

\begin{center}
\includegraphics[width=8cm,height=10cm,angle=270,clip]{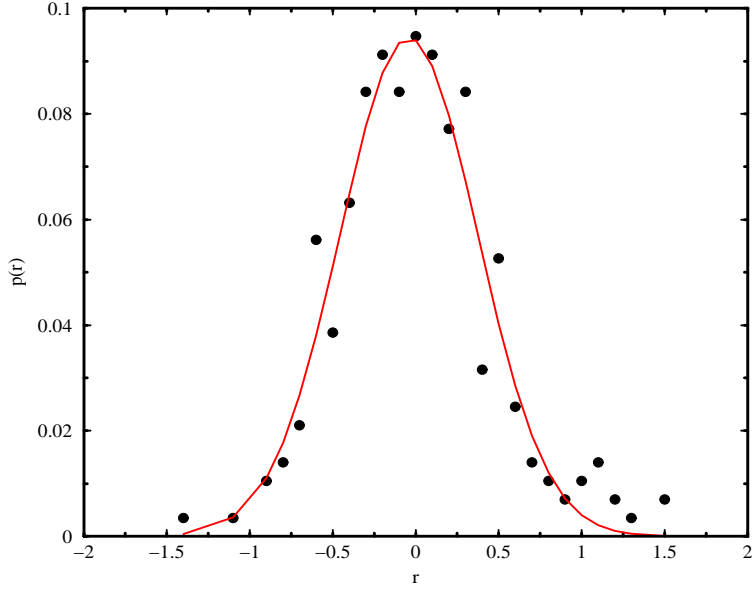}
\end{center}
\caption{The probability distribution $p(r)$ for increments of the number
of the firms bankrupted versus $r=\log n_i - \log n_{i-1}$. 
The solid line is a Gaussian fit.}
\label{f3}
\end{figure}

\begin{figure}[t]

\begin{center}
\includegraphics[width=8cm,height=10cm,angle=270,clip]{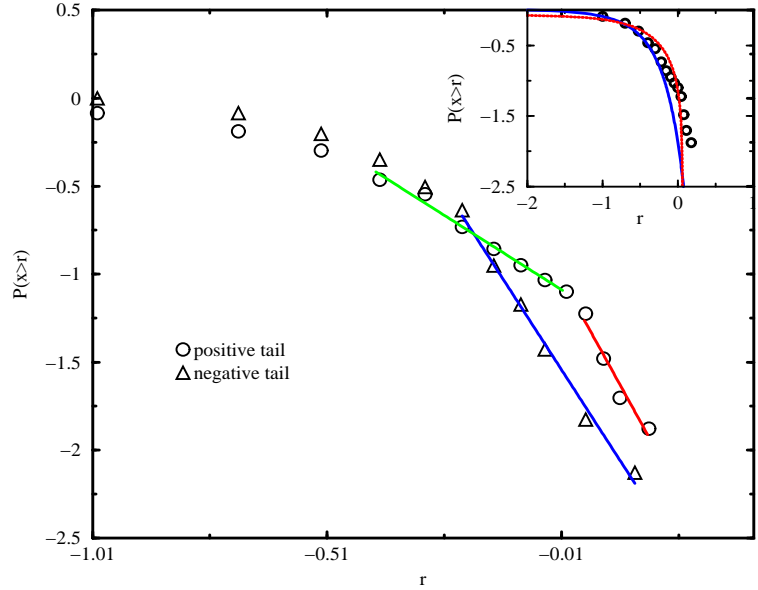}
\end{center}
\caption{The log-log plot of the cumulative probability distribution $P(x>r)$
for change of the number of the firms bankrupted versus the return $r$
for the positive tail($\circ$) and the negative tail($\triangle$). 
The solid lines are the least-square fit with  slopes $-1.69(9)$, $-4.8(8)$ 
for the positive tail and
$-4.1(2)$ for the negative tail. Inset, the solid line is a Gaussian fit
of the cumulative probability distribution for the positive tail. The dotted line
is an exponential fit $p(r)\sim exp(-\alpha r)$. 
The long tail deviates from the Gaussian function.}
\label{f4}
\end{figure}

We also consider  increments of the number of firms bankrupted defined
by 
\begin{equation}
r=\log n_i - \log n_{i-1}.
\end{equation}
The increments $r$ is similar to the return in the analysis of the financial
time series such as stock index\cite{SO03}.
Fig. 3 presents the probability distribution $p(r)$ 
for the increments of the number of the firms bankrupted.
The probability distribution $p(r)$ is well fitted by a Gaussian 
function in the central parts and deviates from the Gaussian at
the tail parts.  However, our observations are limited results
due to the small number of the data sets.

Fig. 4 presents the log-log plot of the cumulative probability distribution
$P(x>r)$ as a function of increment $r$. We observe
the power law, $P(x>r) \sim r^{-\beta}$ for  the positive 
 and negative tails.
For the positive tail we observe two scaling regions with the slopes
$\beta=1.69(9)$ and $\beta=4.8(8)$. 
In the inset of Fig. 4 we observe that the cumulative probability
distribution deviates from the Gaussian fit at large increments.
For the negative tail we observe a single scaling region 
with the slope $\beta=4.1(2)$. 
The scaling exponents are compared to the scaling behaviors
for the increment of the {\em daily volume} in Tokyo stock exchange. Kaijoji and Nuki
reported the exponent $\beta=0.94$ (positive tail) and $\beta=1.02$ (negative 
tail) for increments of volume in Tokyo stock exchange\cite{KN04}.

Let's consider the independence of the successive values $n_i , n_{i+1}$.
We change the variable as $y_i = \ln n_i$. If the pdf of $n_i$ is
$p(n) \sim n^{-(1+\alpha)}$, the pdf of the variable $y$ is given by
$p(y) \sim \exp(-\alpha y)$ by the transformation of the probability
distribution. If successive values of $n$ are uncorrelated, i.e. 
$<y_i y_{i+1}>=0$, the pdf of $r_i = y_i - y_{i+1}$ is simply obtained
by a convolution.
We obtain the pdf of the varialbe $r$ as
$p(r) \sim \exp(-\alpha r) / 2 \alpha$.
In the inset of Fig. 4 we present the pdf, $p(r) \sim \exp(-\alpha r)$
with $\alpha=0.91$. At the small values of $r$ the data are well fitted
by the exponential distribution function. However, in the large value of
$r$ the data deviate from the exponential distribution. We conclude that
there are long range correlations at the parts of the power-law tail.
Long range correlations are an origin of the power-law in the probability
distribution of firms bankrupted.

Fujiwara propsed a model for the dynamics of balance and sheets of a bank
and firms\cite{FU04}. He observed a power law for the probability 
distribution of debts bankrupted. However, his model is not involved 
to the network relations between firms.
Aleksiejuk et al. propsed a model of collective bank bankruptcies\cite{AH01}.
In their model banks are located on a lattice. Banks with surpluses tend to
invest their money, whereas banks with shortfalls borrow money
from the nearest neighbor bank. Above three dimension
the distribution of avalanches for bank bankrupted follows the power
law, $p(s) \sim s^{-\tau}$ with $\tau=1.5$ where $s$ is
the size of avalanche. 

The bankruptcies are a complicated phenomena in economics.
Failure of a firm is caused by the internal and external factors
of the firms. The effects of a bankruptcy of a firm propagate through 
the network of creditor-debtor. The creditors are perturbed by the firm bankrupted.
The bankruptcy induces some chain reactions
to the market and economic systems. If the effects of the bankruptcy are 
weak, the market absorbs the perturbations and recovers from the damages.
However, if the effects are strong, secondary bankruptcies are induced.
The chain reactions of the bankruptcy are created. 
The number of firms bankrupted in the chain reactions is similar to
the number of toppled sites in avalanche of the sand pile\cite{Bak96}.
The herding dynamics is a reason for the scaling 
behavior in the bankruptcies.

We consider the scaling properties of the distribution of fluctuations
in the number of firms bankrupted in Korea. 
We observe the power law distribution of
the number of firms bankrupted. The Pareto exponent is close to 
unity. This Pareto exponent is similar ot Fujiwara's result for
the distribution of debt of bankrupted firms. We also observe the
power law for increments of the number of firms bankrupted. 
The distribution of increments has the different scaling exponents for 
positive and negative tail. 

The bankruptcy of a firm influences to many firms through 
creditor-debtor network. Such avalanche propagates through
the creditor-debtor network. The group dynamics of the firms
induces the self-organized criticality in the bankruptcy.

\section*{Acknowledgments}
The present research has been conducted by the Research Grant
of Kwangwoon University in 2006.



\begin{thebibliography}{}

\bibitem{SO03} D. Sornette,  "Critical Phenomena in Natural Science", 
Springer, Heidelberg, 2003.
\bibitem{ZI49} G. K. Zipf, "Human Behavior and the Principle of Least Effort",
Addison-Wesley, Reading, 1949. 
\bibitem{GR44} B. Gutenberg and R. F. Richter, Bulletin of the Seismological Society 
of America {34} (1944) 185.
\bibitem{GE94} T. Gehrels, "Hazards due to Comets and Asteroids", University 
of Arizona, Tucson, 1994. 
\bibitem{LH91} E. T. Lu and R. J. Hamilton, Astrophysical Journal
  {390} (1991) 89.
\bibitem{RT98} D. C. Roberts and D. L. Turcotte, Fractals  {6}
(1998) 351.
\bibitem{CF95} R. A. K. Cox, J. M. Felton and K. C. Chung, Journal of Cultural Economics
 {19} (1995) 333.
\bibitem{WY22} J. W. Willis and G. U. Yule, Nature  {109} (1922) 177.
\bibitem{FU04} Y. Fujiwara, Physica A {337} (2004) 219.
\bibitem{OT99} K. Okuyama, M. Takayasu, and H. Takayasu, Physica A {269}
(1999) 125.
\bibitem{AS00} H. Aoyama, W. Souma, Y. Nagahara, M. P. Okazaki, H. Takayasu, and
M. Takayasu, Fractal {8} (2000) 293.
\bibitem{IS05} A. Ishikawa, Physica A {349} (2005) 597.
\bibitem{PA04} D. O. Pushkin and H. Aref, Physica A  {336} (2004) 571.
\bibitem{Lee1} K. E. Lee and J. W. Lee, J. Korean Phys. Soc. {44} (2004) 668.
\bibitem{Lee2} J. W. Lee, K. E. Lee, and P. A. Rikvold, J. Korean Phys. Soc. {48} 
S123(2006).
\bibitem{Lee3} J. W. Lee, K. E. Lee, and P. A. Rikvold, Physica A {364} (2006) 355.
\bibitem{RA} http,//www.rating.co.kr.
\bibitem{KN04} T. Kaizoji and M. Nuki, Fractals {12} (2004) 49.
\bibitem{Bak96} P. Bak, "How Nature Works, the science of self-organized criticality",
Springer, New York, 1996.
\bibitem{AH01} A. Aleksiejuk, J. A. Holyst, and G. Kossinets, cond-mat/0111586.

\end{thebibliography}
\end{document}